\documentclass[adp,a4paper,fleqn%
]{w-art}
\usepackage{times,cite,w-thm}
\theoremstyle{plain}

\theoremstyle{definition}

\usepackage[]{graphicx}

\def\bk{{\mathbf k}}

\def\bq{{\mathbf q}}

\def\b0{{\mathbf 0}}

\def\cS{{\cal S}}

\def\up{\uparrow}
\def\down{\downarrow}

\def\alf{\alpha}
\def\eps{\epsilon}

\def\Gam{\Gamma}

\def\Lam{\Lambda}

\def\sg{\sigma}

\def\psib{\bar\psi}

\def\sgn{{\rm sgn}}

\begin{document}
\DOIsuffix{theDOIsuffix}
\Volume{16}
\Month{01}
\Year{2007}
\pagespan{1}{}
\Receiveddate{XXXX}
\Reviseddate{XXXX}
\Accepteddate{XXXX}
\Dateposted{XXXX}



\title[Anomalous criticality near semimetal-to-superfluid transition]
{Anomalous criticality near semimetal-to-superfluid quantum phase
 transition in a two-dimensional Dirac cone model}


\author[B. Obert]{Benjamin Obert\inst{1}}
\address[\inst{1}]{Max-Planck-Institute for Solid State Research,
 Heisenbergstr.\ 1, D-70569 Stuttgart, Germany}
\author[S. Takei]{So Takei\inst{2}}
\address[\inst{2}]{Department of Physics,
 The University of Maryland College Park, MD 20742, USA}
\author[W. Metzner]{Walter Metzner\inst{1,}
\footnote{Corresponding author, E-mail w.metzner@fkf.mpg.de}}
\begin{abstract}
 We analyze the scaling behavior at and near a quantum critical
 point separating a semimetallic from a superfluid phase.
 To this end we compute the renormalization group flow for a model
 of attractively interacting electrons with a linear dispersion
 around a single Dirac point.
 We study both ground state and finite temperature properties.
 In two dimensions, the electrons and the order parameter fluctuations 
 exhibit power-law scaling with anomalous scaling dimensions.
 The quasi-particle weight and the Fermi velocity vanish at the 
 quantum critical point.
 The order parameter correlation length turns out to be infinite
 everywhere in the semimetallic ground state.
\end{abstract}
\maketitle                   







\section{Introduction}

Quantum phase transitions in interacting electron systems are
traditionally described by an effective order parameter theory,
which was pioneered by Hertz \cite{hertz76} and Millis 
\cite{millis93}.
In that approach, an order parameter field is introduced via
a Hubbard-Stratonovich transformation and the electrons are
subsequently integrated out. The resulting effective action for
the order parameter is then truncated at quartic order and
analyzed by standard scaling and renormalization group (RG) 
techniques.

However, more recent studies revealed that the Hertz-Millis
approach is often not applicable, especially in low dimensional
systems \cite{belitz05,loehneysen07}.
For electron systems with a Fermi surface the electronic excitation 
spectrum is gapless. As a consequence, integrating out the electrons
may lead to singular interactions in the effective order parameter
action, which cannot be approximated by a local quartic term.
Therefore it is better to keep the electronic degrees of freedom in
the theory, treating them on equal footing with the bosonic order
parameter field.
Several coupled boson-fermion systems exhibiting quantum criticality
have been analyzed in the last decade by various methods
\cite{vojta00,abanov00,abanov03,rech06,kaul08,huh08}.

Recently, a Dirac cone model describing attractively interacting
electrons with a linear energy-momen\-tum dispersion was introduced
to model a continuous quantum phase transition from a semimetal 
to a superfluid \cite{strack10}.
The scaling behavior at the quantum critical point (QCP) was studied
by coupled boson-fermion flow equations derived within the 
functional RG framework.
It was shown that electrons and bosons acquire anomalous scaling
dimensions in dimensions $d<3$, implying non-Fermi liquid behavior
and non-Gaussian order parameter fluctuations.

In this work we extend the analysis of the Dirac cone model in 
various directions, with a focus on the two-dimensional case.
First, we allow for a renormalization of the Fermi velocity of the
electrons, which was omitted in Ref.~\cite{strack10}, but indeed 
turns out to be important.
Second, we study the behavior upon approaching the QCP from the
semimetallic phase at zero temperature. 
While the pairing susceptibility exhibits the expected power-law
scaling, we find that the correlation length is infinite everywhere
in the semimetallic phase.
Finally, we compute the scaling behavior of the susceptibility and
the correlation length in the finite temperature quantum critical 
regime.

The paper is structured as follows. 
In Sec.~2 we define the Dirac cone model and the corresponding
action.
The derivation of the flow equations is described in Sec.~3, and
results are presented in Sec.~4.
We conclude with a short summary in Sec.~5.


\section{Dirac cone model}

We consider a model of electrons with a linear dispersion relation
$\eps_{\bk\alf} = \alf v_f |\bk|$, with $\alf = \pm 1$, corresponding
to two ``Dirac cones'' with positive ($\alf=1$) and negative 
($\alf=-1$) energy.
The chemical potential is chosen as $\mu = 0$, such that in
the absence of interactions states with negative energy are 
filled, while states with positive energy are empty. The
Fermi surface thus consists of only one point, the ``Dirac 
point'' at $\bk = \b0$, where the two Dirac cones touch.
The action of the interacting system with a local attractive
interaction $U < 0$ is given by \cite{strack10}
\begin{eqnarray}
  \cS[\psi,\psib]
  &=& \int_{k\alf\sg}
  \psib_{k\alf\sg} (-ik_{0} + \eps_{\bk\alf}) \, \psi_{k\alf\sg}
  + U \! \int_{k\alf} \int_{k'\alf'} \! \int_q 
  \psib_{-k,\alf\down} \psib_{k+q,\alf\up}
  \psi_{k'+q,\alf'\up} \psi_{-k',\alf'\down} \nonumber \\
  &+& \int_{k\alf\sg} 
  m_{\alf} \, \psib_{k\alf\sg} \psi_{k\alf\sg} \, ,
  \label{bare_action}
\end{eqnarray}
where $\psi$ and $\psib$ are fermionic fields. 
The variables $k = (k_0,\bk)$ and $q = (q_0,\bq)$ collect Matsubara 
frequencies and momenta, and we use the short-hand notation
$\int_k = T \sum_{k_0} \int \frac{d^d \bk}{(2\pi)^d} \,$
for momentum integrals and frequency sums;
$\int_{k\alf}$ includes also the sum over the band index $\alf$
and $\int_{k\alf\sg}$ includes in addition the spin sum over 
$\sg = \, \up,\down \,$.
Momentum integrations are restricted by the ultraviolet cutoff 
$v_f |\bk| < \Lambda_0$.

In Ref.~\cite{strack10} it was tacitly assumed that the interaction
does not shift the upper and lower Dirac cone with respect to each
other. To compensate for self-energy contributions which in fact do 
generate such a shift, we have added a fermionic mass term with a
$U$-dependent mass $m_{\alf}$ to the action $\cS[\psi,\psib]$.
This term is tuned such that the Dirac cones touch each other at
$\bk = 0$ for any $U$.

The kinetic energy in Eq.~(\ref{bare_action}) is a toy version 
of the dispersion for electrons moving on a honeycomb lattice 
as in graphene, where the momentum dependence is entangled with a
pseudospin degree of freedom related to the two-atom structure
of the unit cell \cite{castroneto09}.
Note that the kinetic energy and the interaction in 
Eq.~(\ref{bare_action}) are both diagonal in the spin indices.
By contrast, in Dirac fermion models describing surface states of 
certain three-dimensional topological insulators the spin orientation
is correlated with the momentum \cite{fu07,moore07}.
We are not aware of a physical realization of the model
Eq.~(\ref{bare_action}) in a real material. The model was designed
to analyze the quantum phase transition between a semimetal and a
superfluid in the simplest possible setting.
Although the model (\ref{bare_action}) is reminiscent of the Gross-Neveu 
model \cite{gross74}, it is not equivalent to it. In particular, 
for the Gross-Neveu model there is no choice of a spinor basis 
in which the kinetic and potential energies are both spin-diagonal.

The attractive interaction favors spin singlet pairing \cite{strack10}. 
Therefore, we decouple the interaction in the s-wave spin-singlet 
pairing channel by introducing a complex bosonic Hubbard-Stratonovich 
field $\phi$ conjugate to the bilinear composite of fermionic fields
$U \int_{k\alf} \psi_{k+q,\alf\up} \psi_{-k,\alf\down} \,$.
This yields a functional integral over $\psi$, $\psib$ and $\phi$
with the fermion-boson action
\begin{eqnarray}
 \cS[\psi,\psib,\phi]
  &=& \int_{k\alf\sg} 
  \psib_{k\alf\sg} (-ik_0 + \eps_{\bk\alf} + m_{\alf}) \, 
  \psi_{k\alf\sg}
  - \int_q \phi^*_q \frac{1}{U} \, \phi_q
  \nonumber\\
  &+& \int_{k\alf} \int_q \left(
  \psib_{-k,\alf\down} \psib_{k+q,\alf\up} \,
  \phi_q +
  \psi_{k+q,\alf\up} \psi_{-k,\alf\down} \,
  \phi^*_q \right) \, , \nonumber\\ 
  \label{finalmodel}
\end{eqnarray}
where $\phi^*$ is the complex conjugate of $\phi$. 
The boson mass $\delta = - 1/U > 0$ is the control parameter for 
the quantum phase transition.
In mean-field theory a continuous transition between the semimetallic
and a superfluid phase occurs at the quantum critical point
$U_{qc}^{\rm MF} = - 2\pi v_f^2/\Lam_0$ in two dimensions
\cite{strack10}.
For technical reasons explained in Sec.~3, we will supplement the 
bosonic part of the action by adding a term of the form
$\int_q \phi_q^* (Z_b q_0^2 + A_b \bq^2) \phi_q$, which regularizes 
the flow at high scales without influencing the low-energy properties 
of the system.
The extra term corresponds to a replacement of the local interaction 
$U$ by a $q$-dependent interaction 
$U(q) = U/[1 - U (Z_b q_0^2 + A_b \bq^2)]$, which decreases at
large momenta and frequencies.
From now on we set $v_f = 1$.


\section{Renormalization group}
\label{sec:RG}

Our aim is to derive scaling properties of the electrons and the 
order parameter fluctuations near the quantum phase transition.
To this end we derive flow equations for the scale-dependent 
effective action $\Gamma^{\Lambda}[\psi,\bar{\psi},\phi]$
within the functional RG framework for fermionic and bosonic 
degrees of freedom \cite{berges02,baier04,schuetz05,strack08}.
Starting from the bare fermion-boson action
$\Gamma^{\Lambda=\Lambda_0}[\psi,\bar{\psi},\phi] =
 \cS[\psi,\bar{\psi},\phi]$
in Eq.~(\ref{finalmodel}), fermionic and bosonic 
fluctuations are integrated simultaneously, proceeding from 
higher to lower scales as parametrized by the continuous 
flow parameter $\Lambda$. 
In the infrared limit $\Lambda \rightarrow 0$, the fully
renormalized effective action $\Gamma[\psi,\bar{\psi},\phi]$ is 
obtained.
The flow of $\Gamma^{\Lambda}$ is governed by the exact functional 
flow equation \cite{berges02}
\begin{eqnarray}
\frac{d}{d\Lam} \Gam^{\Lam}[\psi,\psib,\phi] =
 {\rm Str} \, \frac{\partial_{\Lam} R^{\Lam}}
 {\Gam^{(2) \Lam}[\psi,\psib,\phi] + R^{\Lam}} \; ,
\label{floweq_exact}
\end{eqnarray}
where $\Gam^{(2) \Lam}$ denotes the second functional derivative 
with respect to the fields and $R^{\Lam}$ is the infrared regulator
(to be specified below). The supertrace (Str) traces over 
all indices, with an additional minus sign for fermionic contractions. 

\subsection{Truncation}

The functional flow equation Eq.~(\ref{floweq_exact}) cannot be 
solved exactly.
We therefore truncate the effective action with the objective to 
capture the essential renormalization effects.
Our ansatz for $\Gam^{\Lam}$ is a slight generalization of the 
truncation used in Ref.~\cite{strack10} of the following form
\begin{equation}
 \Gam^{\Lam} = 
 \Gam_{\psib\psi}^{\Lam} + \Gam_{\phi^*\phi}^{\Lam} +
 \Gam_{|\phi|^4}^{\Lam} + \Gam_{\psi^2\phi^*}^{\Lam} \; ,
\label{ansatz}
\end{equation}
where
\begin{eqnarray}
 \Gam_{\psib\psi}^{\Lam} &=& \int_{k\alf\sg}
 \psib_{k\alf\sg} 
 (- i Z_f^{\Lam} k_0 + A_f^{\Lam} \eps_{\bk\alf} + m_{\alf}^{\Lam}) \, 
 \psi_{k\alf\sg} \; , 
 \label{Gam1} \\
 \Gam_{\phi^*\phi}^{\Lam} &=&
 \int_q \phi^*_q \, 
 \left( Z_b^{\Lam} q_0^2 + A_b^{\Lam} \bq^2 + \delta^{\Lam} \right) 
 \phi_q \; ,
 \label{Gam2} \\
 \Gam_{|\phi|^4}^{\Lam} &=&
 \frac{u^{\Lam}}{8} \int_{q,q',p} 
 \phi^*_{q+p} \phi^*_{q'-p} \phi_{q'} \phi_q \; ,
 \label{Gam3} \\
 \Gam_{\psi^2\phi^*}^{\Lam} &=&
 g^{\Lam} \int_{k\alf} \int_q \left(
 \psib_{-k,\alf\down} \psib_{k+q,\alf\up} \,
 \phi_q +
 \psi_{k+q,\alf\up} \psi_{-k,\alf\down} \,
 \phi^*_q \right) \, .
 \label{Gam4}
\end{eqnarray}
The momentum and frequency dependence of $\Gam_{\phi^*\phi}^{\Lam}$,
and also the bosonic interaction $\Gam_{|\phi|^4}^{\Lam}$, are
generated by fermionic fluctuations.
The fermion-boson vertex $\Gam_{\psi^2\phi^*}^{\Lam}$ is actually not 
renormalized within our truncation. 
The usual one-loop vertex correction, which is formally of order 
$g^3$, vanishes in the normal phase due to particle conservation
\cite{strack08}.
Hence, the coupling $g$ remains invariant at its bare value $g=1$ 
in the course of the flow.

In Ref.~\cite{strack10} a restricted version of the ansatz 
Eq.~(\ref{ansatz}) with $A_f^{\Lam} = Z_f^{\Lam}$ and 
$A_b^{\Lam} = Z_b^{\Lam}$ was used, since it was assumed that 
frequency and momentum dependences renormalize similarly.
However, a closer inspection reveals that this is not the case.
In particular, it turns out that one-loop contributions to the
flow of $A_f^{\Lam}$ cancel, while $Z_f^{\Lam}$ flows to infinity 
at the QCP.
This asymmetry between momentum and frequency scaling generates 
also a significant difference between $A_b^{\Lam}$ and $Z_b^{\Lam}$.

The initial conditions for the fermionic renormalization factors
are $Z_f^{\Lam_0} = A_f^{\Lam_0} = 1$. The initial condition for 
the bosonic mass is $\delta^{\Lam_0} = - 1/U$, and the quartic 
bosonic interaction $u$ is initially zero.
The initial conditions for $Z_b$ and $A_b$ corresponding to the 
bare action in Eq.~(\ref{finalmodel}) are $Z_b^{\Lam_0} = 
A_b^{\Lam_0} = 0$. 
However, starting the flow with $Z_b^{\Lam_0} = A_b^{\Lam_0} = 0$ 
leads to very large transient anomalous dimensions at the initial 
stage of the flow (for $\Lam$ near $\Lam_0$), which complicates 
the analysis in a (high energy) regime which is physically not 
interesting. The qualitative behavior of the low energy flow 
($\Lam \ll \Lam_0$) and the critical exponents do not depend on 
the initial values of $Z_b$ and $A_b$. 
We therefore add a term $\int_q \phi_q^* (q_0^2 + \bq^2) \, \phi_q$ 
to the bare action, corresponding to initial values 
$Z_b^{\Lam_0} = A_b^{\Lam_0} = 1$.
This term regularizes the model by suppressing the interaction for 
large momentum and energy transfers.

As regulators in the flow equation (\ref{floweq_exact}) we choose 
momentum dependent Litim functions \cite{litim01}, supplemented by
a mass shift for the fermions,
\begin{eqnarray}
 R_{f\alf}^{\Lam}(\bk) &=& 
 A_f \left[-\Lam \, \text{sgn}(\eps_{\bk\alf}) + 
 \eps_{\bk\alf} \right] \theta(\Lam - |\eps_{\bk\alf}|) +
 \delta m_{\alf}^{\Lam} 
 \; , \nonumber \\
 R_b^{\Lam}(\bq) &=& 
 A_b \left( - \Lam^2 + \bq^2 \right) 
 \theta\left( \Lam^2 - \bq^2 \right) \; ,
\label{cutoffs}
\end{eqnarray}
where $\delta m_{\alf}^{\Lam}$ is chosen such that it cancels
$m_{\alf}^{\Lam}$ in Eq.~(\ref{Gam1}) at each scale $\Lam$.
Note that we have set $v_f = 1$, such that $\Lam$ is a common
momentum cutoff for fermions and bosons.
Adding the regulator functions to the quadratic terms in the
effective action $\Gam^{\Lam}$ yields the inverse of the
regularized propagators, which thus have the form
\begin{equation}
 G_{f\alf}^{\Lam}(k) = \left[
 iZ_f^{\Lam} k_0 - A_f^{\Lam} \eps_{\bk\alf} - m_{\alf}^{\Lam}
 + R_{f\alf}^{\Lam}(\bk) \right]^{-1} =
 \frac{1}{i Z_f^{\Lam} k_0 - A_f^{\Lam} \sgn(\eps_{\bk\alf})
 \max(\Lam,|\eps_{\bk\alf}|)} \; , 
\label{G_f} 
\end{equation}
\begin{equation}
G_b^{\Lam}(q) = \left [
 - Z_b^{\Lam} q_0^2 - A_b^{\Lam} \bq^2 - \delta^{\Lam} 
 + R_b^{\Lam}(\bq) \right]^{-1} =
 - \frac{1}{Z_b^{\Lam} q_0^2 + A_b^{\Lam} \max(\Lam^2,\bq^2) + 
 \delta^{\Lam}} \; .
\label{G_b}
\end{equation}

Symmetry breaking in interacting Fermi systems is often studied
by extending the model to an arbitrary number of fermion flavors
$N_f$, and expanding in the parameter $1/N_f$. 
Our truncation captures the leading contributions for large $N_f$.
The low energy behavior is captured correctly also to leading 
order in $\eps$, where $\eps = 3-d$ is the deviation from the 
critical spatial dimension $d_c=3$, below which anomalous scaling 
sets in.

\subsection{Flow equations}

The flow equations are obtained by inserting the ansatz 
Eq.~(\ref{ansatz}) for $\Gam^{\Lam}$ into the exact functional
flow equation Eq.~(\ref{floweq_exact}) and comparing coefficients.
For a concise formulation, we use the following short-hand 
notation for a cutoff derivative and loop integration:
\begin{eqnarray}
 {\int_k}' = 
 T \sum_{k_0} \int \frac{d^dk}{\left(2\pi\right)^{d}}
 \sum_{s=b,f} \left( - \partial_{\Lam} R_s^{\Lambda} \right)
 \partial_{R_s^{\Lambda}} \; .
\label{intprime}
\end{eqnarray}
The scale-derivatives of the regulators read
\begin{eqnarray}
 \partial_{\Lam} R_{f\alf}^{\Lam}(\bk) &=& 
 -A_f^{\Lam} \, \text{sgn}(\eps_{\bk\alf}) \,
 \theta(\Lam - |\eps_{\bk\alf}|) \; , \nonumber \\
 \partial_{\Lam} R_b^{\Lam}(\bq) &=& 
 - 2 A_b^{\Lam} \Lam \, \theta\left(\Lam^2 - \bq^2 \right) \, ,
\end{eqnarray}
where terms proportional to $\partial_{\Lam} A_f^{\Lam}$ and 
$\partial_{\Lam} A_b^{\Lam}$ are neglected (as usual, see 
\cite{berges02}). The contribution from the mass shift,
$\partial_{\Lam} \delta m_{\alf}^{\Lam}$, is also discarded. 
It is formally of higher order (in a loop expansion) than the 
terms kept, and it does not affect the qualitative behavior.
Note that the cutoff derivative in Eq.~(\ref{intprime}) acts
only on the explicit cutoff dependence introduced via the 
regulator functions.

We thus obtain the following equations for the flow of parameters 
in our ansatz for $\Gam^{\Lam}$:
\begin{eqnarray}
\partial_{\Lam} Z_f^{\Lam} &=& 
 (g^{\Lam})^2 {\int_q}' \left. \frac{\partial}{i\partial k_0} \,
 G_{f\alf}^{\Lam}(q-k) \, G_b^{\Lam}(q) \, \right|_{k=0} \; ,
\label{flow_Zf} \\
 \partial_{\Lam} A_f^{\Lam} &=& 0 \; ,
\label{flow_Af} \\
 \partial_{\Lam}\delta^{\Lam} &=& 
 (g^{\Lam})^2 \int_{k\alf}' 
 G_{f\alf}^{\Lam}(k) \, G_{f\alf}^{\Lam}(-k) + 
 \frac{u^{\Lam}}{2} {\int_q}' G_b^{\Lam}(q) \; ,
\label{flow_delta} \\
 \partial_{\Lam}Z_b^{\Lam} &=& 
 \frac{1}{2} \frac{\partial^2}{\partial q_0^2} \, 
 \left. (g^{\Lam})^2 \int_{k\alf}' 
 G_{f\alf}^{\Lam}(k+q) \, G_{f\alf}^{\Lam}(-k) \, \right|_{q=0} \; ,
\label{flow_Zb} \\
 \partial_{\Lam}A_b^{\Lam} &=&
 \frac{1}{2} \frac{\partial^2}{\partial q_1^2} \, 
 \left. (g^{\Lam})^2 \int_{k\alf}' 
 G_{f\alf}^{\Lam}(k+q) \, G_{f\alf}^{\Lam}(-k) \, \right|_{q=0} \; ,
\label{flow_Ab} \\
 \partial_{\Lam} u^{\Lam} &=& - 4 (g^{\Lam})^4 \int_{k\alf}' 
 \big[ G_{f\alf}^{\Lam}(-k) \big]^2 \big[ G_{f\alf}^{\Lam}(k) \big]^2 +
 \frac{5}{4} (u^{\Lam})^2 {\int_q}' \big[ G_b^{\Lam}(q) \big]^2 \; ,
\label{flow_u} \\
 \partial_{\Lam} g^{\Lam} &=& 0 \; .
\label{flow_g}
\end{eqnarray}
The flow equations for $Z_f$, $\delta$, $Z_b$, $u$, and $g$ are the
same as in Ref.~\cite{strack10}. The momentum derivative in the
flow equation for $A_b$ is with respect to the first (or any other)
component of $\bq$.
All frequency sums and momentum integrations in the above flow 
equations can be performed analytically, both at zero and finite
temperature. 

Explicit $\Lam$-dependences in the flow equations can be 
absorbed by using rescaled variables
\begin{equation}
 \tilde\delta^{\Lam} = \frac{\delta^{\Lam}}{\Lam^2 A_b^{\Lam}} 
 \; , \quad
 \tilde g^{\Lam} =
 \frac{g^{\Lam}}{\Lam^{\frac{3-d}{2}} 
 \sqrt{Z_f^{\Lam} A_f^{\Lam} A_b^{\Lam}}} \; , \quad
 \tilde u^{\Lam} = 
 \frac{u^{\Lam}}{\Lam^{3-d} \sqrt{Z_b^{\Lam} (A_b^{\Lam})^3}} \; .
\label{variables}
\end{equation}
At $T>0$ one also has to use rescaled temperatures
$\tilde T_b^{\Lam} = T \frac{\sqrt{Z_b^{\Lam}/A_b^{\Lam}}}{\Lam}$ 
and $\tilde T_f^{\Lam} = T \frac{Z_f^{\Lam}}{A_f^{\Lam} \Lam}$ 
to absorb $\Lam$.
Anomalous dimensions are defined as usual by logarithmic 
derivatives of the renormalization factors
\begin{equation}
 \eta_b^A = - \frac{d\log A_b^{\Lam}}{d\log\Lam} \; , \quad
 \eta_b^Z = - \frac{d\log Z_b^{\Lam}}{d\log\Lam} \; , \quad
 \eta_f^A = - \frac{d\log A_f^{\Lam}}{d\log\Lam} \; , \quad
 \eta_f^Z = - \frac{d\log Z_f^{\Lam}}{d\log\Lam} \; .
\label{etas}
\end{equation}
Note that $\eta_f^A = 0$, since $A_f^{\Lam}$ does not flow.

 
\section{Results}
\label{sec:results}

We now discuss the scaling behavior as obtained from a solution
of the flow equations, focussing mostly on the two-dimensional
case. Anomalous scaling dimensions occur in dimensions $d<3$ 
\cite{strack10}. We first discuss the ground state, including 
the quantum critical point, and then finite temperatures.
Numerical results depending on the ultraviolet cutoff $\Lam_0$
will be presented for the choice $\Lam_0 = 1$.

\subsection{Quantum critical point}
\label{sec:qcp}

To reach the quantum critical point one has to tune the bare
interacting to a special value $U_{qc}$ such that the bosonic 
mass $\delta^{\Lam}$ scales to zero for $\Lam \to 0$.
In two dimensions we find $U_{qc} \approx - 15.646$ for 
$\Lam_0 = 1$, which is about a factor $2.5$ larger than the 
mean-field value.
For $U = U_{qc}$ the rescaled variables defined in 
Eq.~(\ref{variables}) scale to a non-Gaussian fixed point,
with finite anomalous dimensions in any dimension $d < 3$.
Since $g^{\Lam}$ does not flow at all, the scale invariance 
of $\tilde g^{\Lam}$ at the fixed point leads to a simple 
relation between the anomalous dimensions,
\begin{equation}
 \eta_b^A + \eta_f^Z + \eta_f^A = 3 - d \; .
\label{rel1}
\end{equation}
Furthermore, since the flow of $Z_b$ and $A_b$ is determined 
entirely by a convolution of two fermionic propagators, the 
differences of anomalous dimensions for frequency and momentum
scaling of fermions and bosons are linked by a simple condition, 
which can be expressed as
\begin{equation}
 \eta_b^Z - \eta_b^A = 2(\eta_f^Z - \eta_f^A) \; .
\label{rel2}
\end{equation}
Due to $\eta_f^A = 0$ the above relations reduce to
$\eta_b^A + \eta_f^Z = 3 - d$ and $\eta_b^Z - \eta_b^A = 2\eta_f^Z$.
Solving the fixed point equations we obtain the numerical values
$\eta_b^A \approx 0.75$, $\eta_b^Z \approx 1.25$, and 
$\eta_f^Z \approx 0.25$ in two dimensions.
Hence, at the quantum critical point the order parameter exhibits
non-Gaussian critical fluctuations with different anomalous scaling 
dimensions for momentum and frequency dependences. Furthermore, the
fermionic quasiparticle weight ($\propto Z_f^{-1}$) vanishes, which
implies non-Fermi liquid behavior. Since $A_f$ remains finite, the
Fermi velocity also vanishes at the quantum critical point.
This last point was missed in Ref.~\cite{strack10}.
Due to the different anomalous dimensions for momentum and frequency
scaling, the dynamical exponent $z$ acquires an anomalous dimension, 
too. 
In the bare action $\cS$ one has $z_0 = 1$ for bosons and fermions.
At the quantum critical point, we find
\begin{equation}
 z_f = 1 + \eta_f^Z - \eta_f^A = 
 z_b = 1 + \frac{\eta_b^Z - \eta_b^A}{2} \approx 1.25 \; .
\end{equation}
The equality between $z_b$ and $z_f$ follows from Eq.~(\ref{rel2}).

\subsection{Semimetallic ground state}
\label{sec:semimetal}

For $|U| < |U_{qc}|$, the bosonic mass $\delta^{\Lam}$ saturates at 
a finite value for $\Lam \to 0$, corresponding to a finite pairing
susceptibility $\chi = \lim_{\Lam \to 0} (\delta^{\Lam})^{-1}$.
The fermionic $Z$-factor also saturates, such that $\eta_f^Z \to 0$.
Hence, fermionic quasiparticles survive in the semimetallic state.
However, $A_b^{\Lam}$ and $Z_b^{\Lam}$ do not saturate, but rather 
diverge as $\Lam^{-1}$, such that $\eta_b^A,\eta_b^Z \to 1$.
This is illustrated in Fig.~1, where the anomalous dimensions are
plotted as a function of $\Lam$ for a choice of $U$ close to the
QCP. The QCP scaling is seen at intermediate scales, before the
anomalous dimensions saturate at the asymptotic values 
$\eta_f^Z = 0$ and $\eta_b^A = \eta_b^Z = 1$ for $\Lam \to 0$.
\begin{figure}
\includegraphics[width=8cm]{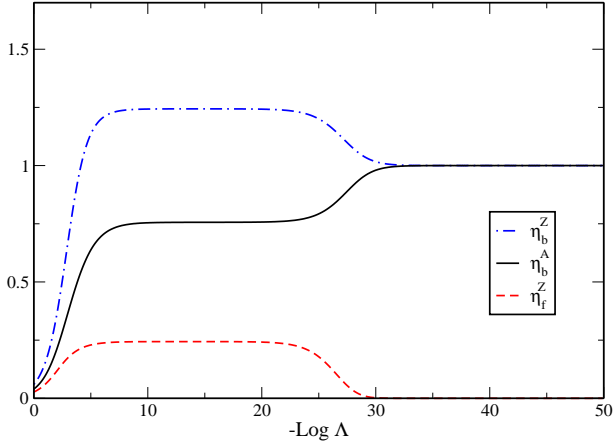}
\caption{Flow of anomalous dimensions as a function of $\Lam$
 in the semimetallic ground state for a choice of $U$ close to 
 the QCP.}
\label{fig:1}
\end{figure}
A finite anomalous dimension away from the critical point is 
surprising at first sight. However, it can be explained quite 
easily. 
An explicit calculation shows that the leading small momentum 
and small frequency dependence of the fermionic particle-particle 
bubble is {\em linear} in two dimensions, 
as long as the propagators have a finite quasi-particle weight. 
In presence of an infrared cutoff this linear behavior is replaced
by a quadratic behavior (as in our ansatz), but the prefactors of
the quadratic terms diverge linearly in the limit $\Lam \to 0$,
reflecting thus the true asymptotic behavior.

The divergences of $A_b^{\Lam}$ and $Z_b^{\Lam}$ imply that the 
correlation length and correlation time of pairing fluctuations 
are always infinite in the semimetallic ground state, not only 
at the QCP. 
This is consistent with the observation that the linear momentum
and frequency dependence of the particle-particle bubble leads
to a power-law decay of its Fourier transform at long space or
time distances, instead of the usual exponential decay.
The divergent correlation length suggests that the entire
semimetallic phase is in some sense ``quantum critical''. 
This point of view has indeed been adopted in theories of 
interaction effects in graphene, where the particle-hole 
symmetric (Dirac) point is interpreted as a QCP separating the 
electron-doped from the hole-doped Fermi liquid. 
Scaling concepts could then be used to compute thermodynamic 
\cite{sheehy07} and transport \cite{fritz08} properties of
graphene near the Dirac point.

The pairing susceptibility $\chi$ is generically finite in the
semimetallic ground state and diverges upon approaching the QCP.
From a numerical solution of the flow equations in two dimensions
we have obtained the power-law
\begin{equation}
 \chi(U) \propto (|U_{qc}| - |U|)^{-\gamma_0} \; , \quad
 \mbox{with} \; \gamma_0 \approx 1.725 \; .
\end{equation}

\subsection{Temperature scaling}
\label{sec:t-scaling}

We now present results for $U = U_{qc}$ and $T > 0$, that is, 
we approach the QCP as a function of temperature.
At finite temperature the fermionic propagator is cut off by
temperature itself, since fermionic Matsubara frequencies are
bounded by $|k_0| \geq \pi T$, and the bosonic propagator is
regularized by the finite bosonic mass $\delta$. 
Hence, the flow of all unscaled variables saturates for 
$\Lam \to 0$.
Power-laws are obtained for these saturated variables as a
function of temperature.
In particular,
\begin{equation}
 A_b \propto T^{-\bar\eta_b^A} \; , \quad
 Z_b \propto T^{-\bar\eta_b^Z} \; , \quad
 Z_f \propto T^{-\bar\eta_f^Z} \; , \quad
\end{equation}
with $\bar\eta_b^A \approx 0.60$, $\bar\eta_b^Z \approx 1.00$,
and $\bar\eta_f^Z \approx 0.20$ in two dimensions.

In Fig.~2 we show the temperature dependence of the 
susceptibility $\chi$ and the correlation length $\xi$, as 
obtained from a numerical solution of the flow equations at 
various temperatures in two dimensions. The susceptibility
is given by the inverse bosonic mass $\delta$ at the end of
the flow ($\Lam \to 0$), the correlation length by
$\xi = \sqrt{Z_b/\delta}$.
\begin{figure}
\includegraphics[width=9cm]{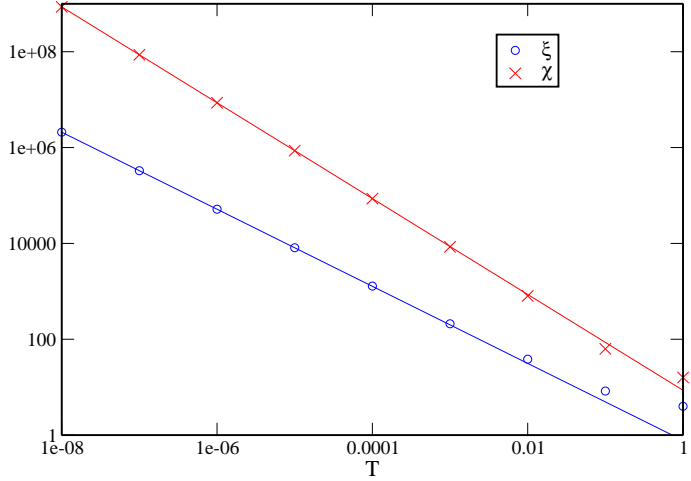}
\caption{Temperature dependence of the pairing susceptibility 
 $\chi$ and the correlation length $\xi$ in a double-logarithmic 
 plot for $U = U_{qc}$. At low temperatures the calculated points
 lie on straight lines, corresponding to power-law behavior.}
\label{fig:2}
\end{figure}
Both quantities obey power-laws at low temperatures, namely
\begin{eqnarray}
 \chi(T) \, \propto & \! T^{-\gamma} \; , \quad
 & \mbox{with} \; \gamma \approx 1.00 \; , \\
 \xi(T) \, \propto & \! T^{-\nu} \; , \quad
 & \mbox{with} \; \nu \approx 0.80 \; .
\end{eqnarray}
Note that we use the letters $\gamma$ and $\nu$ for the exponents 
by applying the classical definition near a thermal phase transition 
$\chi \propto (T - T_c)^{-\gamma}$ and $\xi \propto (T - T_c)^{-\nu}$ 
to the present situation where $T_c = 0$.
The correlation length exponent obeys $\nu = z_b^{-1}$, which 
corresponds to a $T^{-1}$ scaling of the correlation time $\xi_{\tau}$
in accordance with general scaling arguments for quantum phase 
transitions.
The exponents $\gamma$ and $\nu$ obey the classical scaling relation 
$\gamma = (2- \eta_b^A) \nu$.


\section{Conclusion}
\label{sec:conclusion}

We have analyzed the critical properties near a quantum phase transition
between a semimetallic and a superfluid phase in a two-dimensional model
of attractively interacting electrons with a Dirac cone dispersion,
correcting and extending a previous work \cite{strack10}.
We have studied coupled flow equations for the fermionic degrees of 
freedom and the bosonic fluctuations associated with the superfluid 
order parameter.
Both fermions and bosons acquire anomalous scaling dimensions at the
QCP, corresponding to non-Fermi liquid behavior and non-Gaussian pairing 
fluctuations.
Allowing for distinct renormalization factors for momentum and frequency 
scaling, we have found that they differ substantially at the QCP. 
In particular, the Fermi velocity vanishes.
We have also analyzed the semimetallic ground state away from the QCP
in more detail than previously, finding that the correlation length 
for pairing fluctuations is always infinite, not only at the QCP.
Finally, we have studied the scaling behavior upon approaching the
QCP as a function of temperature.
The susceptibility and the correlation length obey power-laws in
temperature, as expected, and the corresponding critical exponents
obey the classical scaling relation.


\begin{acknowledgement}
This work is dedicated to Dieter Vollhardt on the occasion of his
60th birthday, to honor his influential research on correlated electrons 
and superfluidity, and to acknowledge his valuable support of young 
scientists at early stages of their career.
We thank H.~Gies, P.~Jakubczyk, V.~Juricic, S.~Sachdev, P.~Strack,
and O.~Vafek for helpful discussions.
We also gratefully acknowledge support by the DFG research group FOR 723.
\end{acknowledgement}


\end{document}